\documentclass[prd,twocolumn,amsmath,amssymb,floatfix,superscriptaddress,nofootinbib,preprintnumbers]{revtex4-2}
\pdfoutput=1
\usepackage{color}
\usepackage{graphicx}
\usepackage{tabularx}
\usepackage[dvipsnames, table]{xcolor}
\usepackage{xspace}
\usepackage[justification=centerlast, format=plain, labelfont=bf]{caption}
\usepackage{subcaption}
\usepackage{float}
\usepackage{verbatim}
\usepackage{amsmath}\usepackage{graphicx}
\usepackage{tabularx}
\usepackage{amssymb}
\usepackage{url}
\usepackage{bbold}
\usepackage{slashed}
\usepackage{array}
\usepackage{multirow}
\usepackage{threeparttable}
\usepackage{paralist}
\usepackage{xspace}
\usepackage{upgreek}
\usepackage{lipsum}
\usepackage[normalem]{ulem}

\usepackage[colorlinks]{hyperref}

\newcolumntype{C}{>{\centering\arraybackslash}X}

\newcommand{\be}{\begin{equation}}
  \newcommand{\ee}{\end{equation}}
    \newcommand{\ba}{\begin{align}}
  \newcommand{\ea}{\end{align}}

\begin{document}

\title{Towards an Optimal Estimation of Cosmological Parameters with the Wavelet Scattering Transform}

\author{Georgios Valogiannis}
 \email{gvalogiannis@g.harvard.edu}
\author{Cora Dvorkin}
 \email{cdvorkin@g.harvard.edu}
\affiliation{
 Department of Physics, Harvard University, Cambridge, MA, 02138, USA\\
}

\begin{abstract}
Optimal extraction of the non-Gaussian information encoded in the Large-Scale Structure (LSS) of the universe lies at the forefront of modern precision cosmology. We propose achieving this task through the use of the Wavelet Scattering Transform (WST), which subjects an input field to a layer of non-linear transformations that are sensitive to non-Gaussianity in spatial density distributions. In order to assess its applicability in the context of LSS surveys, we apply the WST on the 3D overdensity field obtained by the {\it Quijote} simulations, out of which we extract the Fisher information in 6 cosmological parameters. WST delivers a large improvement in the marginalized errors on all parameters, ranging between $1.2-4\times$ tighter than the corresponding ones obtained from the regular 3D cold dark matter + baryon power spectrum, as well as a $50 \%$ improvement over the neutrino mass constraint given by the marked power spectrum. Through this first application on 3D cosmological fields, we demonstrate the great promise held by this novel statistic and set the stage for its future application to actual galaxy observations. 
\end{abstract}

\maketitle

\section{Introduction}

The current and next of stage of cosmological surveys, such as the Dark Energy Spectroscopic Instrument (DESI) \citep{Levi:2013gra}, the Vera Rubin Observatory Legacy Survey of Space and Time (LSST) \citep{Abell:2009aa,Abate:2012za}, Euclid \citep{Laureijs:2011gra} or the the {\it Nancy Grace Roman Space Telescope} \citep{Spergel:2013tha}, will map out the 3-dimensional (3D) distribution of galaxies in the large-scale structure (LSS) of the universe at unprecedented levels of accuracy. Given that the latter acts as an ideal window into the fundamental processes that have shaped its formation, we will thus get a unique opportunity to shed light on the nature of the accelerated expansion of the universe \citep{Copeland:2006wr}, dark matter \citep{LSSTDarkMatterGroup:2019mwo}, massive neutrinos \citep{LESGOURGUES2006307,Dvorkin:2019jgs}, the properties of gravity at large scales \citep{Ishak:2018his,doi:10.1146/annurev-astro-091918-104423, Alam:2020jdv}, and the physics of the early universe \citep{Chen_2016}. 

Despite the tremendous promise held by these ambitious observational endeavors, a full exploitation of the information encoded in the 3D cosmic web is highly non-trivial. From a theoretical standpoint, one of the simplest and meaningful compressions of the cosmological information, the power spectrum (2-point function) of density fluctuations, fails to capture all relevant information. This is the case due to the non-linear process of gravitational instability, responsible for structure formation, which has caused the emergence of higher-order correlations \citep{PhysRevLett.108.071301}; in other words, cosmic density fields have become highly non-Gaussian. Despite the notable theoretical and numerical progress on understanding and extracting higher-order clustering statistics made in the last decade \citep{BERNARDEAU20021,Hahn_2020, Hahn_2021}, the substantial computational cost associated with such evaluations, which rises sharply with increasing order in the correlation function, makes including them in standard cosmological analyses still infeasible.

As a consequence of this major challenge, a wide range of alternative techniques has been proposed in the literature, aiming for a shorter path to accessing the non-Gaussian information component of the LSS. Approaches of this kind include, but are not limited to, using proxy estimators \citep{PhysRevD.91.043530,Dizgah_2020}, exploiting the greater information content in the unvirializied regions of the LSS (cosmic voids) \citep{Pisani:2019cvo,Massara_2015, 10.1093/mnras/stz1944, 10.1093/mnras/stv777, Hamaus_2015, Kreisch:2021xzq}, penalizing the overdensities with non-linear transformations that aim to capture the information that lies beyond the ordinary power spectrum
 \citep{Neyrinck_2009,PhysRevLett.108.071301,PhysRevLett.107.271301,White_2016,PhysRevD.97.023535,PhysRevLett.126.011301}, using 1-point statistics \citep{Uhlemann:2019gni}, the k-nearest neighbor cumulative distribution functions \citep{Banerjee:2020umh}, or the minimum spanning tree \citep{Naidoo:2021dxz}. Even though the exploration of such techniques is a very active area of research, the extent to which they can reliably replace traditional estimators is still an open question. Last but not least, Convolutional Neural Networks (CNNs) \citep{6522407} have recently shown great promise at improving cosmological parameter constraints by extracting information from non-Gaussian structures \citep{PhysRevD.97.103515,Villaescusa_Navarro_2021}, but fall prey to known interpretability issues associated with their application on real data. 

In this work we propose tackling this problem with a novel estimator, in the context of the 3D density field, the Wavelet Scattering Transform (WST), which was first introduced in the context of signal processing \cite{https://doi.org/10.1002/cpa.21413} as an ideal middle-ground between conventional statistics and CNNs \citep{6522407}. The WST consists of a series of nonlinear transformations of an input field, in what can be regarded as a CNN with fixed filters and fixed weights of well-understood properties, that can efficiently and reliably capture non-Gaussian information \citep{https://doi.org/10.1002/cpa.21413,10.1214/14-AOS1276}. Wavelet transforms have been growing increasingly popular in multiple subfields of astrophysics, recently finding successful applications in studies of the interstellar medium \citep{refId0, Saydjari_2021,refdust}, weak lensing (WL) \cite{10.1093/mnras/staa3165,Cheng:2021hdp} and 2D slices of cosmological density fields \citep{PhysRevD.102.103506}. Aiming to fully leverage the information content in realistic LSS observations, as performed by modern galaxy surveys, in this paper we apply the WST on the 3D density field, for the first time in the literature. We then forecast its ability to constrain cosmological parameters, using N-body simulations, demonstrating large improvement over the traditional power spectrum analysis. Our analysis thus sets the scene for a future WST application on the \textit{biased} galaxy density field, which is what LSS surveys observe \citep{DESJACQUES20181}, rather than the underlying matter overdensity field. 

\section{Wavelet Scattering Transform}

In the WST \cite{https://doi.org/10.1002/cpa.21413,6522407} an input field, $I(\vec{x})$, is convolved by a family of localized wavelets generated by performing spatial and angular dilations of a mother wavelet, thus picking out various scales of interest. Followed by a modulus operation and taking the spatial average gives rise to a set of coefficients that are sensitive to the clustering properties of the input field at the oriented scale to which the wavelet is most sensitive. Repeating the process encodes information on higher-order $N$-point correlation functions, with the $N^{th}$ layer including contributions from up to the $2^{N}$-point correlation function \citep{https://doi.org/10.1002/cpa.21413,10.1214/14-AOS1276}. 

As opposed to previous cosmological applications of wavelet techniques \citep{10.1093/mnras/staa3165,Cheng:2021hdp,PhysRevD.102.103506} that restricted their attention on 2D configurations only, in this work we aim to study the information content of the WST applied on the full 3D density field of the LSS. To do so, we will closely follow the corresponding 3D WST implementation of \citep{doi:10.1063/1.5023798}, given by the publicly available {\it KYMATIO} package \citep{2018arXiv181211214A}\footnote{\url{https://www.kymat.io/}},to evaluate the WST coefficients, $S_n$, up to order $n=2$. 

The Wavelet Scattering Transform (WST) coefficients, $S_n$, up to order $n=2$, are given by the following relations:

\begin{align} \label{eq:WSTcoeff:sup}
 S_0 &= \langle |I(\vec{x})|^{q} \rangle, \nonumber \\
 S_1(j_1,l_1) &= \left\langle \left(\sum_{m=-l_1}^{m=l_1}|I(\vec{x}) \ast \psi^{m}_{j_1, l_1}(\vec{x}) |^{2}\right)^{\frac{q}{2}} \right\rangle, \\
 S_2(j_2,j_1,l_1) &= \left\langle \left(\sum_{m=1}^{m=l_1}|U_1(j_1,l_1)(\vec{x}) \ast \psi^{m}_{j_2, l_1}(\vec{x}) |^{2}\right)^{\frac{q}{2}} \right\rangle \nonumber,
\end{align}
based on the particular 3D implementation of Ref. \citep{doi:10.1063/1.5023798}, given by the publicly available {\it KYMATIO} package \citep{2018arXiv181211214A}. In Eq. (\ref{eq:WSTcoeff:sup}), $\langle.\rangle$ indicates spatial averaging, while $U_1(j_1,l_1)(\vec{x})$ denotes the first order convolution of the input field
\begin{equation}\label{u1}
 U_1(j_1,l_1)(\vec{x}) = \left(\sum_{m=-l_1}^{m=l_1}|I(\vec{x}) \ast \psi^{m}_{j_1, l_1}(\vec{x}) |^{2}\right)^{\frac{1}{2}}.
\end{equation}
We note that convolution in Eq. (\ref{u1}) (and all higher orders) is performed in the Fourier space, assuming periodic boundary conditions. Furthermore, $\psi^{m}_{j_1, l_1}$ reflect dilations of the mother wavelets of the form
\begin{equation}\label{solid:sup}
\psi^{m}_{l}(\vec{x}) = \frac{1}{\left(2\pi\right)^{3/2}}e^{-|\vec{x}|^2 /2 \sigma^2}|\vec{x}|^l Y_l^m \left(\frac{\vec{x}}{|\vec{x}|}\right),
\end{equation}
described by the relationship 
\begin{equation}\label{dil:sup}
\psi^{m}_{j_1, l_1}(\vec{x}) = 2^{-3 j_1}\psi^{m_1}_{l_1}(2^{-j_1} \vec{x}),
\end{equation}
where the maximum number of spatial dilations and angular oscillations, $(j_1,l_1)$, are given by $(J,L)$, respectively, and $Y_l^m$ are the familiar Laplacian Spherical Harmonics. Wavelets of the form (\ref{solid:sup}), first used in the context of molecular chemistry \citep{doi:10.1063/1.5023798,10.5555/3295222.3295400} and 3D image processing \citep{10.1007/978-3-642-03798-6_14}, are called ``solid harmonic wavelets", because they consist of a solid harmonic, $|\vec{x}|^l Y_l^m \left(\frac{\vec{x}}{|\vec{x}|}\right)$, multiplied by a Gaussian with a width $\sigma$ (in units of pixels). Even though this particular wavelet form was motivated in a different context, it was further found to perform very well at capturing non-Gaussianities encoded in 3D interstellar medium maps \citep{Saydjari_2021}, as we also report in our cosmological analysis here. Aside from the dimensionality, fundamentally the main difference between this solid harmonic implementation and the 2D WST version applied on WL maps \cite{10.1093/mnras/staa3165,Cheng:2021hdp} is the use of a different wavelet form as a convolution filter, with a ``Morlet" wavelet being the corresponding choice in the latter case\footnote{We also add that in the 2D WST case the modulus is strictly raised to the power of $q=1$.}. Furthermore, in direct analogy to the chemistry investigations mentioned above, one may further optimize the performance of the WST for LSS applications in particular, using wavelets designed to leverage existing symmetries (as for example in Ref. \citep{2021arXiv210411244S}). We plan to explore these issues in upcoming work. 

We also note that, unlike the 2D implementation, the second order coefficients of the 3D WST evaluated from the {\it KYMATIO} package (and based on Eq. \ref{eq:WSTcoeff:sup}) are a function of a single angular scale only (that is, $S_2 \equiv S_2(j_2,j_1,l_1)$ rather than $S_2(j_2,j_1,l_2,l_1)$). From a practical standpoint, this choice reduces the associated computational cost and the dimensionality of the observable vector by sacrificing a part of the angular information (since $l_2=l_1$). It will be interesting to investigate (in future work) the extent to which the addition of the second angular scale improves the performance of the WST, including in the angle-averaged ISO case, which seems to perform worse than using all coefficients with the current configuration. Furthermore, only second order coefficients with $j_2 > j_1$ are evaluated, both in the 2D and 3D cases. This is the case because, after performing the first order wavelet convolution with $J=j_1$, information encoded at scales smaller than the one defined by $j_1$ is significantly suppressed. Indeed, second order coefficients with $j_1 < j_2$ were not found to contribute any further information in the WL applications of Ref. \cite{10.1093/mnras/staa3165,Cheng:2021hdp}. We note, however, that this is not necessarily always the case, as evidenced by \citep{2021arXiv210411244S}. Using equivariant wavelets applied on 2D images, some residual power was found to lie in the coefficients usually discarded\footnote{This can be seen by the interactive Figures in \url{https://faun.rc.fas.harvard.edu/saydjari/EqWS/interactive_corners.html}}. In addition,  \citep{2021arXiv210411244S} did not find a clear divide in the usefulness of the various second order coefficients, as can be seen from Fig. 11 of that work.  Given a maximum scale $J$ and number of angular bins $L$, with the indices $(j,l) \in ([0,..,J-1,J],[0,..,L-1,L])$, the combination of the above choices results in $(L+1)(J+1)$ $1^{st}$ order and $(L+1)(J+1)J/2$ $2^{nd}$ order coefficients, for a total of $S_0+S_1+S_2=1+(L+1)(J^2+3J+2)/2$ coefficients generated up to second order. Coefficients with a greater degree of isotropy, $S_{\rm Iso}$, can be further constructed \cite{10.1093/mnras/staa3165,refId0, Saydjari_2021}, by averaging over all $l$'s for a given $j$ combination, thus reducing the number of $S_1+S_2$ coefficients by a factor of $(L+1)$. For a more sophisticated attempt at preserving isotropy, see \cite{2021arXiv210411244S}. Unless stated otherwise, we hereafter choose the pair of values $J=4$ and $L=4$ as our baseline case, which corresponds to a total of 76 coefficients and 16 isotropic ones. It is worth noting, at this point, that using higher $J$ and $L$ values is perfectly acceptable and will most likely improve the constraints given by the WST. The memory requirement for these computations, however, exceeds the capabilities of the GPUs at hand, so we restrict our attention to the choice of $J=4$ and $L=4$.

\begin{figure}[b]
\centering 
\includegraphics[width=0.49\textwidth]{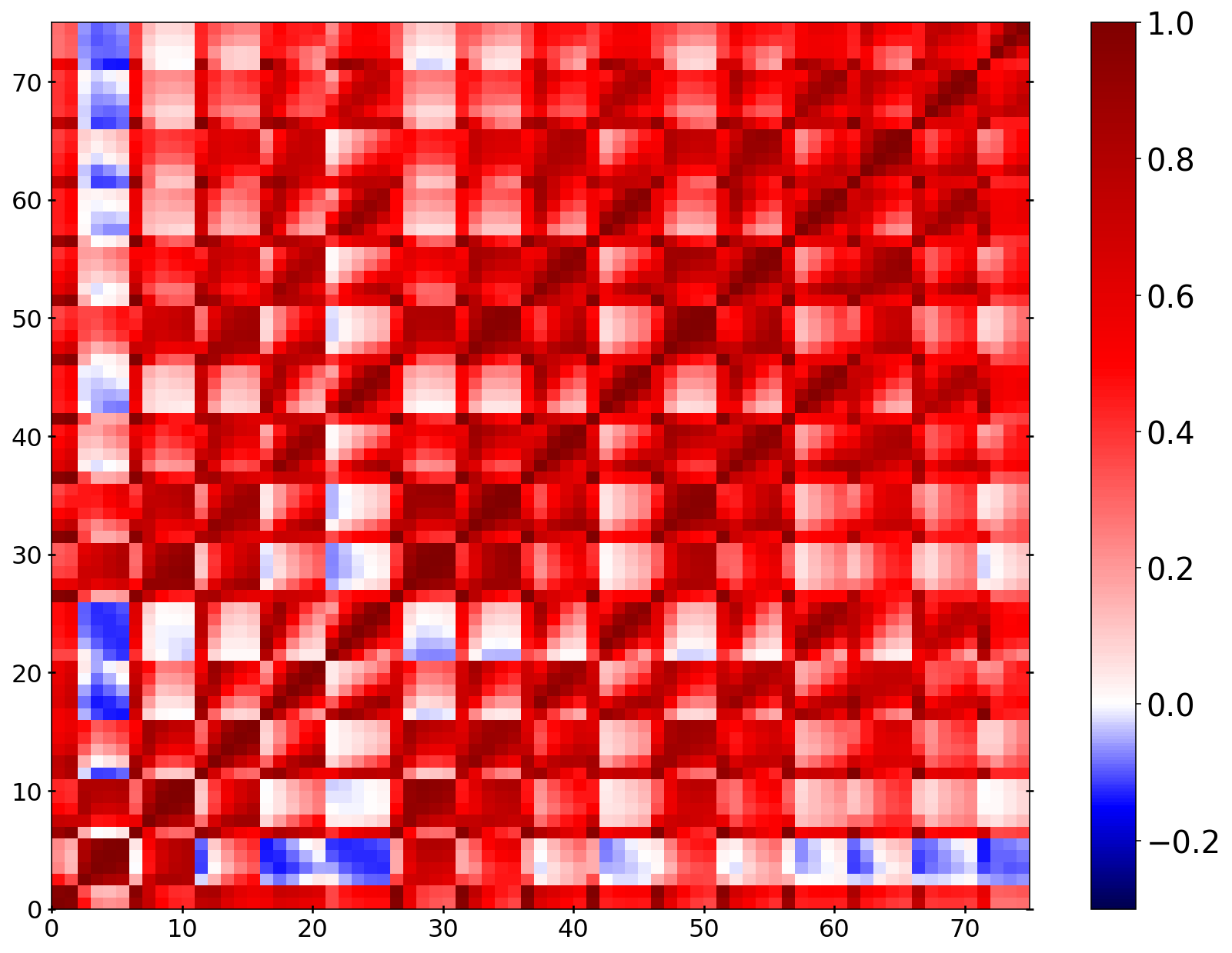}
\caption{\label{fig:epsart} Correlation matrix of all 76 coefficients for the optimal WST case evaluated at the fiducial cosmology. The WST coefficients populate the data vector in order of increasing values of the $j_1$ and $l_1$ indices, with the $l_1$ index varied faster.}
\label{Figapp1}
\end{figure}

Furthermore, the power index $q$, to which the modulus of the convolved field is raised in Eq. \eqref{eq:WSTcoeff:sup}, allows us to explore different density-weighting schemes, with $q=1$ corresponding to the standard choice of using the modulus \citep{https://doi.org/10.1002/cpa.21413}, while values of $q>1$ emphasize on the peaks of the target input field (in our case, cosmological overdensities). More importantly, values of $q<1$ effectively up-weight lower non-zero values (underdensities), providing a measure of sparsity \citep{doi:10.1063/1.5023798}. Given that cosmic voids are known to serve as powerful probes of cosmological information \citep{Pisani:2019cvo}, from massive neutrinos \citep{Massara_2015, 10.1093/mnras/stz1944, PhysRevLett.126.011301} to the properties of gravity \citep{10.1093/mnras/stv777, Hamaus_2015, PhysRevD.97.023535}, we investigate the case of $q<1$ below. Lastly, we consider variations in the wavelet's Gaussian width $\sigma$ in Eq. \eqref{solid:sup}. 

We finally comment on the fact that the WST coefficients are commonly normalized as follows:
 \begin{align} \label{eq:WSTnorm:sup}
 \bar{S}_0 &= \log(S_0), \nonumber\\
 \bar{S}_1 &= \log(S_1/S_0), \\
 \bar{S}_2 &= \log(S_2/S_1), \nonumber 
\end{align}
in the 2D case \citep{10.1214/14-AOS1276,refId0, Saydjari_2021,10.1093/mnras/staa3165,Cheng:2021hdp}. The WL studies in Refs. \cite{10.1093/mnras/staa3165,Cheng:2021hdp}, in particular, found the ``decorrelated" second order coefficients to be particularly efficient at breaking degeneracies in the $\Omega_m$-$\sigma_8$ plane. We did consider this variant in our 3D application (including with and without taking the logarithm of the coefficients), without however finding any noticeable changes in the Fisher results and, as a consequence, we report the outcome from using the bare coefficients, for simplicity. 

\section{N-body Simulations}

We apply the WST relations in Eq. \eqref{eq:WSTcoeff:sup} to the 3D matter overdensity field, 
\begin{equation}\label{eq:deltaef}
\delta_m(\vec{x}) = \frac{\rho_m(\vec{x})}{\bar{\rho}_m} -1,
\end{equation}
obtained from N-body simulations. In particular, we use 3D overdensity fields on a grid of $256^3$ resolution obtained by the {\it Quijote} simulations \citep{Villaescusa_Navarro_2020}\footnote{\url{https://quijote-simulations.readthedocs.io/}}. These are N-body simulations that evolved $512^3$ cold dark matter (CDM) particles on a cubic simulation box of 1.0 Gpc/h side, for a fiducial cosmology that corresponds to the following parameter values: $\Omega_m = 0.3175$, $\Omega_b = 0.049$, $H_0 = 67.11$ km/s/Mpc, $n_s = 0.9624$, $\sigma_8 = 0.834$, sum of the neutrino masses $M_{\nu} = 0.0$ eV and dark energy equation of state $w = -1$. In order to enable parameter forecasting, additional simulations have been performed by varying each of the base parameters in a step-wise fashion, while keeping the rest fixed, with all the associated details summarized in Table 1 of \citep{Villaescusa_Navarro_2020}. We note that the neutrino simulations additionally evolved  $512^3$ neutrino particles, in which case both the total matter overdensity, $\delta_m = \delta_{CDM}+\delta_b+\delta_{\nu}$, and the CDM+baryon combination, $\delta_{cb} = \delta_{CDM}+\delta_b$, were stored. We study the density fields at redshift $z=0$. 

\section{Fisher forecast}

We will use the Fisher information approach to forecast the parameter constraints that we could achieve with a WST analysis. It is important to keep in mind that under this approach we are assuming that the likelihood is Gaussian around the fiducial model. In the context of a weak gravitational lensing analysis \citep{ChengChengSiHao:2021hja}, the distribution of the WST coefficients has been seen to be very close to Gaussian. This is because, as opposed to the higher-point functions, where the multiplication of more random variables increases the skewness of the distribution, in the case of the WST, the modulus of always one random variable does not amplify the tails of the distribution. In fact, as more modes are considered, the central limit theorem Gaussianizes the coefficients in a faster way compared to higher-point functions. Furthermore, down-weighting the high-density regions of the LSS has been found to lead to a greater degree of Gaussianity \citep{Neyrinck_2009,PhysRevLett.108.071301,PhysRevLett.107.271301,PhysRevD.97.023535}, a property also shared by the WST through raising the modulus to powers $q<1$ (and also by the marked power spectrum we introduce in Appendix~\ref{sec:markedPk}.)

Based on the standard Fisher formalism, if $O_m$ is a vector of a set of $m$ independent observables that are functions of $\theta_{\alpha}$ cosmological parameters, the Fisher matrix $F$ is then given by 
\begin{equation}\label{Fisher}
F_{\alpha \beta} = \frac{\partial O_i}{\partial \theta_\alpha} C^{-1}_{ij} \frac{\partial O^T_j}{\partial \theta_\beta}, 
\end{equation}
with $C_{ij}$ the covariance matrix of $O_m$, that can be well-approximated as cosmology-independent \citep{refId0Car,Tegmark_1997}. It can then follow that the marginalized $1\sigma$ error on each of the parameters $\theta_{\alpha}$ can never be smaller than $\sigma_{\alpha} = \sqrt{\left(F^{-1}\right)_{\alpha\alpha}}$.

The WST coefficients obtained from Eq. \eqref{eq:WSTcoeff:sup} are sensitive to the fundamental properties of the underlying density field and can thus serve as a new cosmological observable. In order to quantify its ability to constrain cosmological parameters, we employ the Fisher formalism to estimate the marginalized errors, $\sigma_{\alpha}$, on 6 $\Lambda$CDM parameters $\theta_{\alpha}=\{\Omega_m, \Omega_b, H_0, n_s, \sigma_8, M_{\nu}\}$, with the WST coefficients obtained from the {\it Quijote} suite as an observable. In order to assess the potential improvements by the WST over more conventional statistics, we additionally evaluate and compare against the marginalized constraints obtained by the 3D matter power spectrum, $P(k)$, as well as another recently introduced estimator with similar advantages as the WST, the marked power spectrum $M(k)$ \citep{https://doi.org/10.1002/mana.19841160115} (defined in greater detail in Appendix \ref{sec:markedPk}), as it was considered in Refs. \citep{White_2016, PhysRevD.97.023535, PhysRevLett.126.011301} (the motivation for this comparison will become apparent in the next section). The optimal parameter choices for the $P(k)$ and $M(k)$ where chosen to agree with the corresponding publications that considered them in the context of the {\it Quijote} simulations \citep{Villaescusa_Navarro_2020, PhysRevLett.126.011301}, using 91 k-bins up to $k_{max}=0.58$ h/Mpc, and  picking $p=2$, $\delta_s=0.25$ and $R=10$ Mpc/h for the free parameters of the $M(k)$. 

We make use of the full set of the available 15000 (500) realizations for the numerical evaluation of the covariance matrices (derivatives). The correlation matrix, $C_{ij}/(C_{ii}C_{jj})$, evaluated for the 76 WST coefficients (using the optimal WST settings) at the fiducial cosmology is shown in Fig.~\ref{Figapp1}. In order to make sure we obtain an unbiased estimate of the inverse covariance matrix in Eq. (\ref{Fisher}), we apply the standard Hartlap correction factor \citep{refId22} upon inversion, given by:
\begin{equation}\label{Hartlap}
\hat{C}^{-1}_{ij} = \frac{N_r-D-2}{N_r-1}C_{ij}^{-1}, 
\end{equation}
where $N_r=15000$ is the number of available realizations and $D$ the dimensionality of the observable vector, which is equal to $D=76$ for the WST and $D=91$ for the other two estimators we consider (the power spectrum and marked power spectrum). Before inverting the covariance matrix of the WST coefficients, in addition, we carefully checked that its conditioning number is within acceptable limits for the particular WST configuration and number of realizations used, in order to make sure that the highly correlated features observed in Fig.~\ref{Figapp1} do not compromise the accuracy of our Fisher analysis. We similarly confirmed this to be the case for the corresponding matrices of the power spectrum and marked power spectrum, as well.

\begin{figure}[b]
\includegraphics[width=0.49\textwidth]{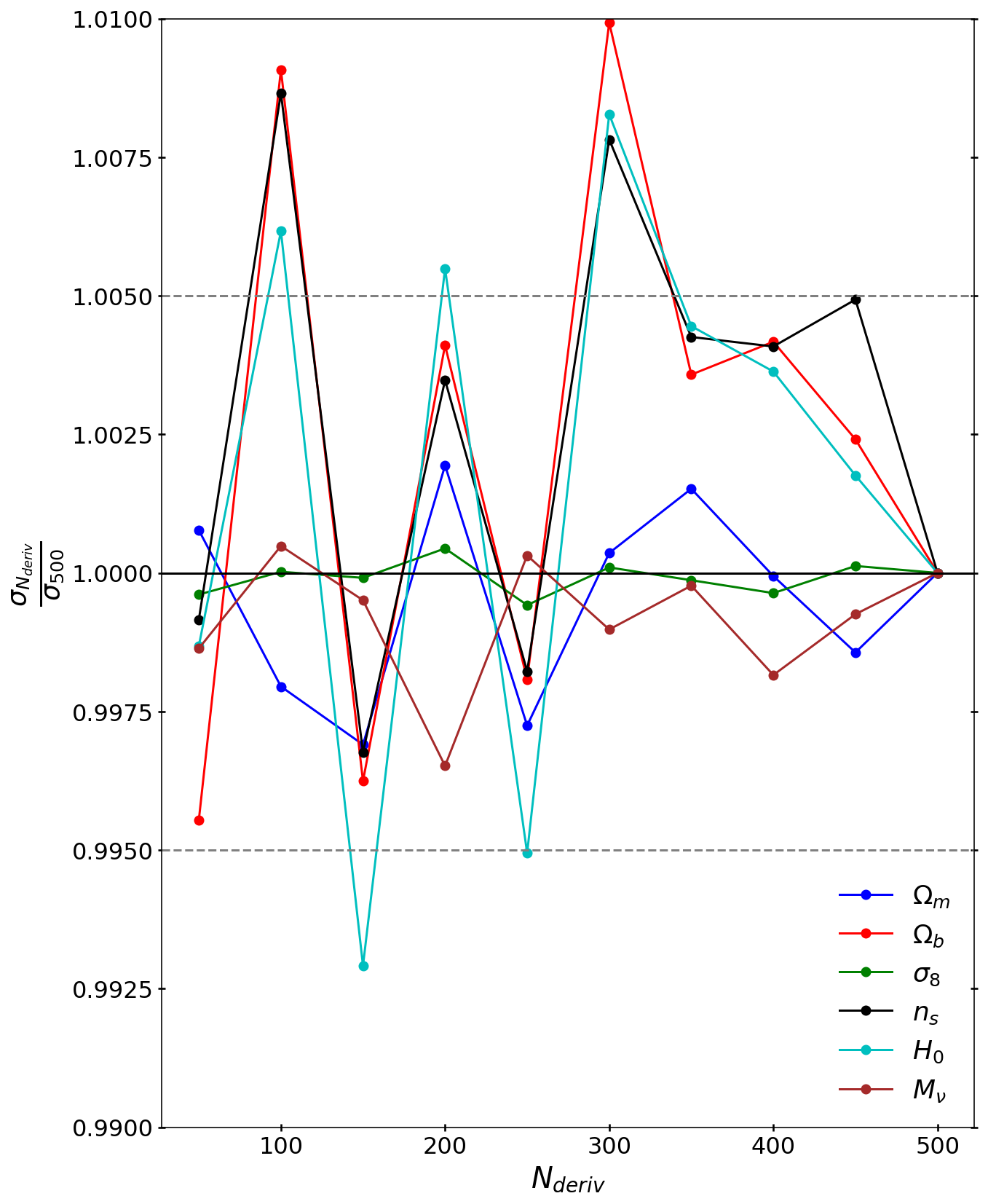}
\caption{The $1\sigma$ errors on the cosmological parameters forecasted by the WST coefficients, $\sigma_{N_{\rm deriv}}$, are plotted as a function of the number of realizations, $N_{\rm deriv}$, used for the numerical evaluation of the derivatives in Eq. (\ref{Fisher}), normalized by the corresponding prediction using all 500 available realizations, $\sigma_{500}$. }
\label{Figapp2}
\end{figure}

\begin{figure*}[b]
\centering 
\includegraphics[width=0.9\textwidth]{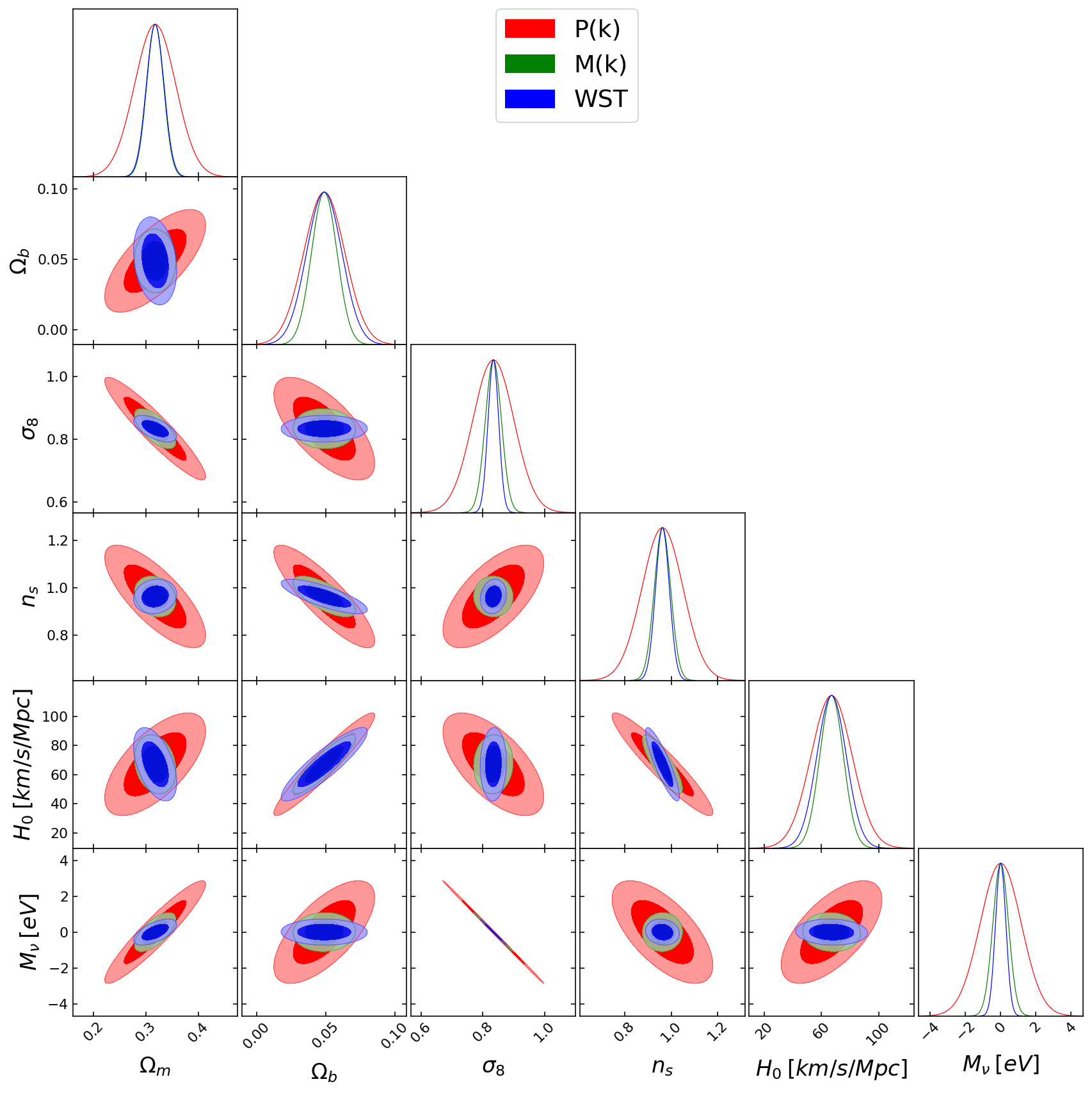}
\caption{\label{fig:app3} Forecasted 1-dimensional and 2-dimensional distribution functions for all 6 cosmological parameters obtained by the power spectrum $P(k)$ (red), the marked power spectrum $M(k)$ (green) and the optimal WST (blue), when working with the density field traced by \textit{CDM+baryons}. The corresponding normalized probability density functions for each parameter are shown along the diagonal.}
\end{figure*}

As far as the numerical evaluation of the derivatives in Eq. (\ref{Fisher}) is concerned, we use all 500 realizations available for each of the parameter steps away from the fiducial. We further add that, in the particular case of the neutrino mass variations, which is a central part of our analysis, $3$ separate steps have been run away from the fiducial of $M_{\nu} = 0.0$ eV, corresponding to values of $M^{+}_{\nu} = 0.10$ eV, $M^{++}_{\nu} = 0.20$ eV and $M^{+++}_{\nu} = 0.40$ eV, using $512^3$ additional neutrino particles and performed with Zel'dovich initial conditions (details explained in Ref. \citep{Villaescusa_Navarro_2020}). We make use of all three neutrino steps $\{M^{+}_{\nu}, M^{++}_{\nu}, M^{+++}_{\nu}\}$ in order to evaluate the derivative with a high order approximation, and against the 500 seeds of the fiducial Zel'dovich set. We have additionally confirmed that our conclusions do not change if we use any of the other lower-order versions for the neutrino derivatives.

The numerical convergence of the marginalized $1\sigma$ errors on all 6 parameters obtained from the WST is plotted as a function of the 500 realizations of the derivatives in Fig. \ref{Figapp2}. We confirm that the variations in the $1\sigma$ predictions (with respect to the results using all 500 seeds) never rise above $1 \%$, even when a very low number of realizations is used, while they converge at the level of $<0.5 \%$ when we use $N_{\rm deriv}\geqslant 350$ seeds. For the error on the sum of the neutrino masses, in particular, for which we report the most impressive improvement over the power spectrum prediction, we note that the convergence is always better than $0.35 \%$. Thanks to the large number of realizations available for the fiducial cosmology, in addition, the forecasted $1\sigma$ errors do not vary by more than $0.1 \%$ when using 10000 out of the total of 15000 realizations to estimate the covariance matrix, for all the parameters. We also find the same levels of convergence for all elements of the Fisher matrix. The results in Fig. \ref{Figapp2} were obtained using the total matter combination, with the convergence behavior being very similar when only working with the matter traced by CDM+baryons. We also confirmed the numerical convergence of the Fisher predictions from the power spectrum and the marked power spectrum, in agreement with their corresponding original Fisher studies presented in Refs. \citep{Villaescusa_Navarro_2020, PhysRevLett.126.011301}.

Last but not least, the forecast for constraints on all 6 cosmological parameters obtained from all 3 observables (and in the `cb' case) is shown in Fig. \ref{fig:app3}, demonstrating the large improvement discussed in the next section.

\section{Results}\label{sec:results}

\begin{figure}[b]
\includegraphics[width=0.49\textwidth]{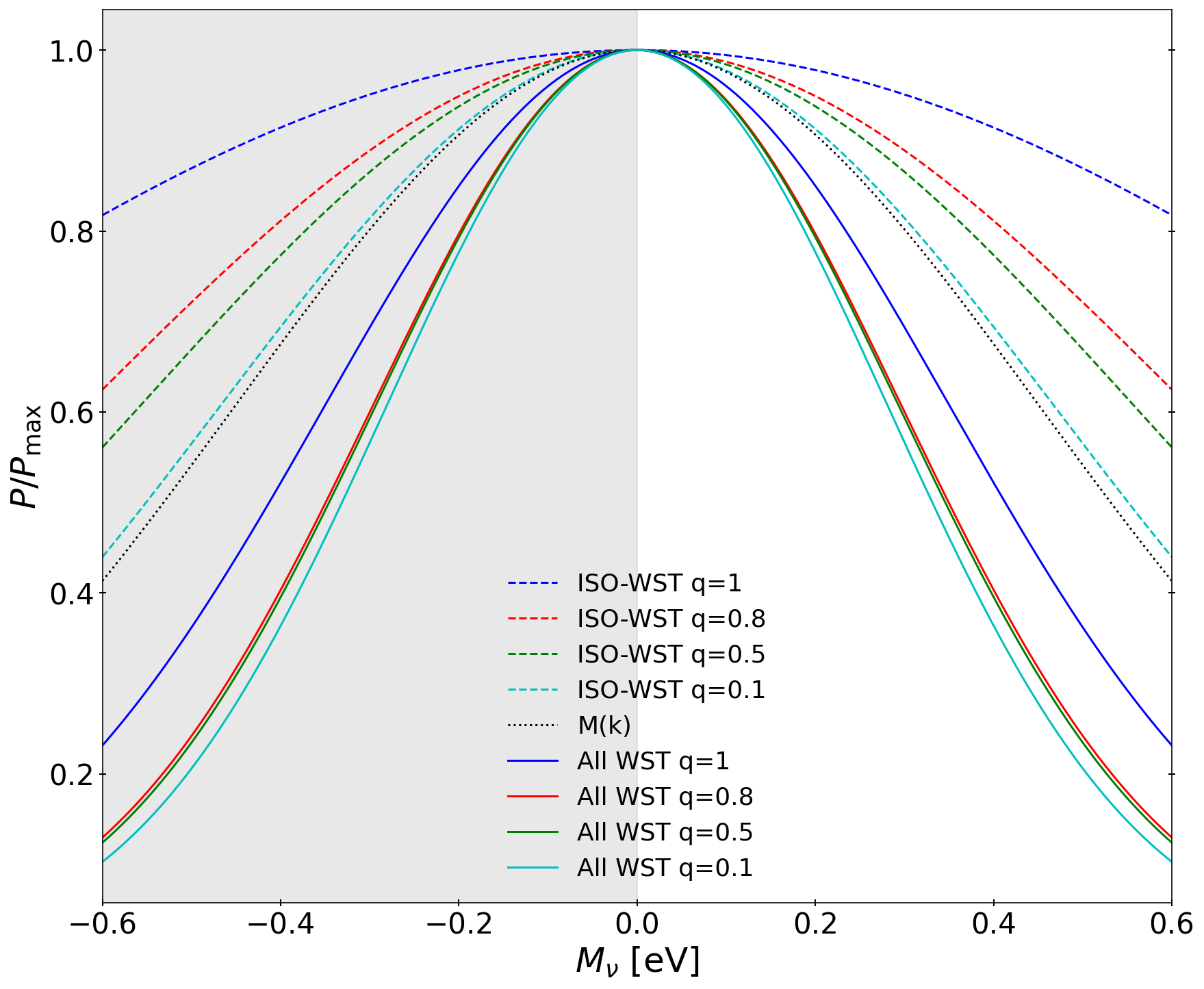}
\caption{\label{fig:1} 1D PDF of the neutrino mass, $M_{\nu}$, as forecasted by using all WST coefficients (solid lines), the isotropic WST coefficients (dashed lines), as the well as the marked power spectrum (black dotted line) as observables, when working with the density given by \textit{CDM+baryons}. Different colors indicate the WST result (with a width $\sigma=0.8$) for varying values of the free parameter $q$ (see Eq. \ref{eq:WSTcoeff:sup}).}
\end{figure}
We begin by determining the optimal choice of the WST free parameters $q$ and $\sigma$ that we will use as the baseline case in this analysis. We vary $q$ in the range $[0.1,1]$ (with a step of $0.1$) and $\sigma$ in $[0.8,1]$ (with a step of $0.05$) and find the pair $\left(q, \sigma\right)=\left(0.8, 0.8\right)$ to be the optimal combination that (jointly) minimizes the average 1-$\sigma$ error for the parameters $\{\Omega_m, \Omega_b, M_{\nu}\}$, which we hereafter adopt. We caution against raising the modulus to very low or high values of $q$, which leads to numerical instabilities in the Fisher estimates. Furthermore, high $q$ values are known to make results susceptible to outliers \citep{PhysRevD.102.103506,ZHANG2021199}. In a similar spirit, and even though lower $\sigma$ values further improve the forecasted results, we do not consider $\sigma<0.8$, to avoid entering the regime where the simulation becomes inaccurate. If we wanted to consider lower values of $\sigma$, that could access scales close to or beyond the resolution limit of the current simulations, a check against the higher resolution Quijote run would be necessary.  In order to illustrate the power of the WST, in Fig.~\ref{fig:1} we show how the marginalized $1\sigma$ error on the neutrino mass, $M_{\nu}$, behaves as a function of the parameter $q$ (when working with the density field traced by CDM+baryons, and for $\sigma=0.8$). The forecasted errors, which are tight in general, monotonically decrease further for lower values of $q$. This behavior is explained by the fact that, as mentioned in the previous section, values of $q<1$ effectively up-weight the significance of low overdensity regions, which are in turn characterized by a known sensitivity to the properties of massive neutrinos \citep{Massara_2015, 10.1093/mnras/stz1944, PhysRevLett.126.011301}, and this is conveniently exploited by this particular weighting scheme. If one wants to solely focus on constraining neutrino masses, lower $q$ values could be chosen, with Fig. \ref{fig:1} as a guide. The numerical convergence of our Fisher errors for the optimal case is better than $1\%$, for all parameters. 

We also briefly comment on the performance of the WST when using the isotropic (``ISO") coefficients. The looser constraints in Fig. \ref{fig:1} are likely attributed to the relatively low value of $J$ we use, which leads to only $16$ coefficients, a problem that can be potentially alleviated by using a higher $J$ value, when more GPU memory becomes available. We plan to further explore this issue in future work, and hereafter only work with all 76 WST coefficients, noting that, even in the ISO case, the constraints to the neutrino mass from Fig. \ref{fig:1} are already $\sim 2 \times$ tighter than when using the regular $P(k)$.

\begin{figure}[b]
\centering 
\includegraphics[width=0.49\textwidth]{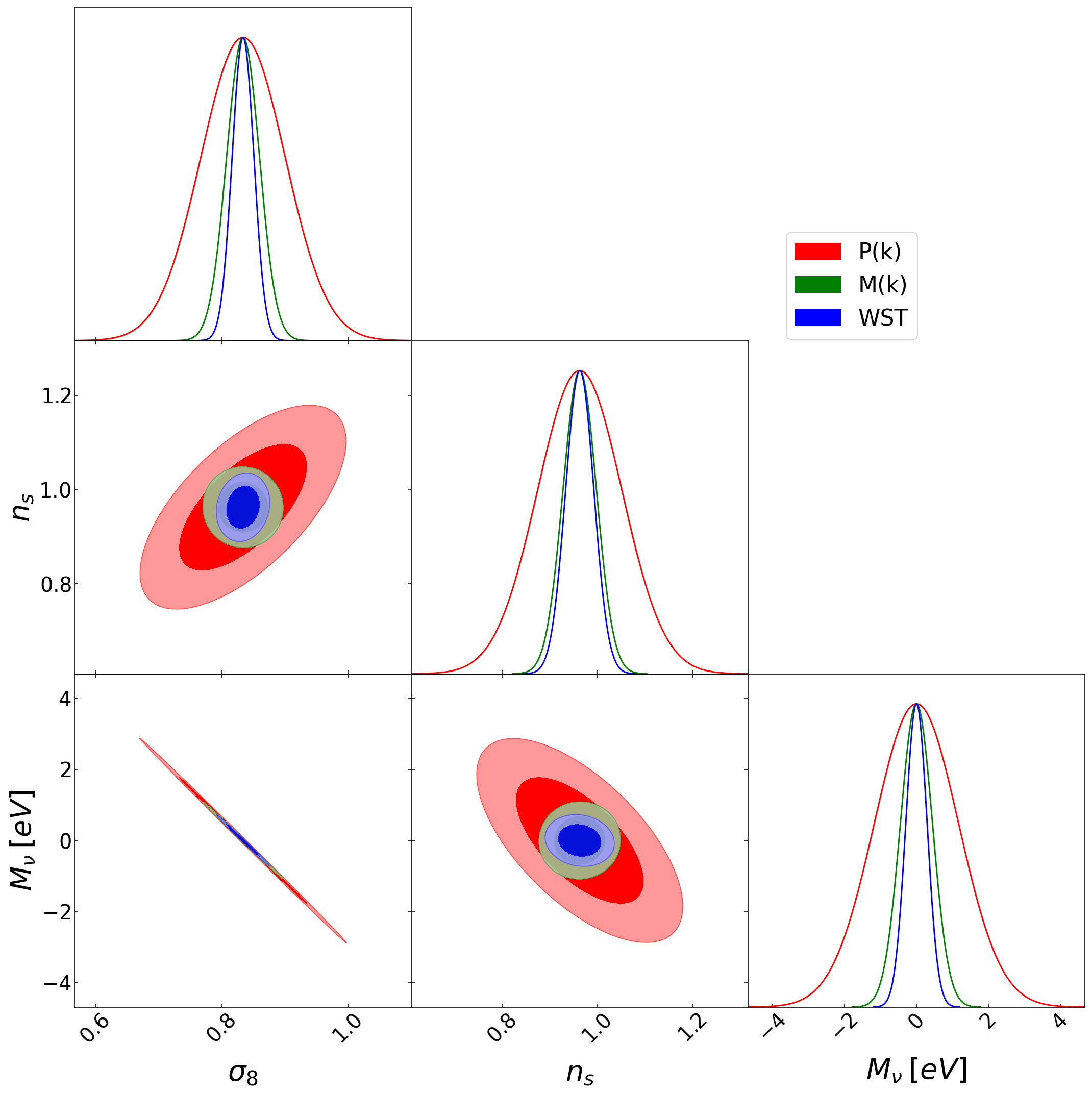}
\caption{\label{fig:2} Forecasted 1$\sigma$ (solid) and 2$\sigma$ (light) constraints to the cosmological parameters $M_{\nu}$, $\sigma_8$, and $n_{s}$ obtained by the power spectrum (red), the marked power spectrum (green) and the optimal WST (blue), when working with the density given by \textit{CDM+baryons}.}
\end{figure}

Having determined the optimal choice of WST coefficients, we proceed to quantify its information content using the Fisher formalism, as explained previously. In Table~\ref{tab1}, we list the marginalized $1\sigma$ errors on each of the 6 cosmological parameters, as obtained by the best case WST, as well as by the other two estimators we consider. We further differentiate between the constraints obtained when using the total matter density field (`m') vs. tracing only the CDM+baryon combination (`cb') in the presence of massive neutrinos. Starting with the neutrino mass, the WST has an impressive performance, predicting $1\sigma$ errors that are $\sim 100\times$ and $\sim 4\times$ tighter than the corresponding ones from the regular $P(k)$-statistics, in the `m' and `cb' cases, respectively. Looking into the rest of the cosmological parameters, and the `m' case, the gain with respect to the $P(k)$ is still substantial, ranging between $\sim 3-15\times$ better. Notably, the predicted errors on $n_s$ and $\sigma_8$ are an order of magnitude tighter than those from the $P(k)$, even though these parameters were not taken into account in the optimization. The $1\sigma$ error ellipses for the 3 parameters demonstrating the greatest improvement, $\{n_s, \sigma_8, M_{\nu}\}$, are shown in Fig.~\ref{fig:2}, using the `cb' field (while the corresponding ellipses for all 6 parameters are shown in Fig. \ref{fig:app3}). The smaller improvement in the neutrino mass constraint (relative to the one coming from $P(k)$) for the `cb' as compared to `m' case is expected, as recently shown in \citep{Bayer:2021kwg}, and indeed the 
`cb' field is directly linked to an observable \citep{Villaescusa_Navarro_2014,Castorina_2014}, as opposed to the 3-dimensional matter field. Nevertheless, the WST does still deliver a substantial decrease by a factor of $1.2-4$ in the forecasted errors, for all parameters, in this case as well.

         \begin{table}[t!]
    	\begin{tabular}{ | p{8.2em} | p{2.6em} |p{2.6em} |p{2.6em}   || p{2.6em} |p{2.6em} |p{2.6em}|}
		\hline
		{Matter type} & \multicolumn{3}{c||}{`m'} & \multicolumn{3}{c|}{`cb'}
		\\ \hline
    	Statistic &  $P(k)$ &M(k) & WST &  $P(k)$ &$M(k)$ & WST 	
		\\ \hline \hline
		$\sigma(\Omega_m)$
		& 0.076 & 0.013 & 0.014 
		& 0.040 & 0.016 & 0.016 
		\\ \hline
		$\sigma(\Omega_b)$
		& 0.033 & 0.010 & 0.012 
		& 0.015 & 0.009 & 0.012 
		\\ \hline
		$\sigma(\sigma_8)$
		& 0.01 & 0.002 &  0.001 
		& 0.067 & 0.026 & 0.017 
		\\ \hline
		$\sigma(n_s)$
		& 0.39 & 0.044 &  0.031 
		& 0.088 & 0.035 & 0.029 
		\\ \hline
		$\sigma(H_0)$ [km/s/Mpc]
		& 40.62 & 9.50 &  10.34 
		& 14.42 & 8.28 & 10.32 
		\\ \hline
		$\sigma(M_\nu$) [eV]
		& 0.72 & 0.016  &  \textbf{0.008} 
		& 1.17 & 0.45 & \textbf{0.29}	 
		\\ \hline
		\end{tabular}
			\caption{Marginalized $1\sigma$ errors on cosmological parameters obtained from $P(k)$, $M(k)$, and optimal WST, when using the total matter field `m', and only CDM+baryons `cb'.
			}
    	\label{tab1}
	\end{table}

The predicted improvement with respect to the neutrino mass estimates exceeds the one recently presented by another similar estimator, the marked power spectrum, in \citep{PhysRevLett.126.011301}, with a ratio of $\sim 1.5\times$ and $\sim 2\times$ tighter in the `cb' and `m' cases, respectively. For the rest of the parameters the 2 estimators have a similar performance, with the WST giving tighter constraints for $\Omega_m$, $n_s$ and $\sigma_8$ and the $M(k)$ doing slightly better for the rest of the parameters (in the `cb' case). 

By performing 2 successive layers of wavelet convolutions to the overdensity field, the WST is picking up information beyond the power spectrum (containing information related up to the 4-point function \citep{https://doi.org/10.1002/cpa.21413,10.1214/14-AOS1276}). It has already been shown that higher-order correlation functions tighten constraints to cosmological parameters (see the case of the bispectrum as an example \citep{Hahn_2020,Hahn_2021}). Additionally, using $q<1$ emphasizes the significance of lower density regions, which are known to carry a wealth of cosmological information \citep{Pisani:2019cvo}.  By realizing these properties in a completely distinct way, the WST can thus serve as a powerful complementary tool for studies of the LSS. One of its great advantages, in particular, is that it seems to be quite immune to the degeneracies between the different cosmological parameters when studying all matter species, in particular with respect to the neutrino mass. The information content from the conventional $P(k)$, on the other hand, tends to saturate at small scales as a result of these degeneracies in the same case (e.g., Fig. 6 of \citep{Villaescusa_Navarro_2020}). We note the parallel to the properties of the marked power spectrum \citep{White_2016, PhysRevLett.126.011301}, which possibly explains the similar performance of the two estimators and our choice to compare them side by side. However, we clarify that this is not quite the case in the `cb' case, for which parameter degeneracies are still present (for example, between $\sigma_8-M_{\nu}$), and as a result we would expect to see less dramatic gains in a future analysis of LSS data, upon marginalizing over the galaxy bias.

Furthermore, we note that the redshift-space bispectrum has been recently also found to significantly increase the information content extracted from the halo \citep{Hahn_2020} and the galaxy \citep{Hahn_2021} density fields, consistent with the expected gain from going beyond the power spectrum. Given, however, that our analysis is instead performed on the underlying matter field and in the real space, these results are not directly comparable to ours and we reserve a detailed 'apples-to-apples' comparison for upcoming work.

We now comment on how our results connect to previous cosmological applications of wavelet techniques. Recently, a 2D version of the WST was shown to greatly improve constraints with respect to $\Omega_m$ and $\sigma_8$ \cite{10.1093/mnras/staa3165} and then also $w$ and $M_{\nu}$ \cite{Cheng:2021hdp}.  Given that these works were focused on a different cosmological probe (WL convergence maps) and for only a subset of the parameters we considered, no direct comparisons can be drawn. In order, however, to better demonstrate the improvement when opting for a full 3D analysis, we briefly quote constraints obtained after applying the 2D WST on the same set of {\it Quijote} simulations. In particular, we obtain 2D projected $256^2$ slices of the original 3D $256^3$ density cubes, by taking the average over the z-axis of each. We then apply the 2D WST for $J=8$ and $L=4$ (as in \cite{10.1093/mnras/staa3165,Cheng:2021hdp}) on the projected overdensity fields, and repeat the Fisher analysis using the new set of generated coefficients as the observable, in the `m' case. The forecasted errors, while more directly related to an observable, are larger than the ones obtained by the 3D matter $P(k)$, for 5 out of the 6 cosmological parameters, by a factor in the range $1.03-7$. This result is not surprising and highlights the need to utilize the information encoded in the full 3D galaxy density field. For the sum of the neutrino masses, however, the 2D WST still gives a tighter constraint than the matter $P(k)$, by about $15 \%$, demonstrating its great sensitivity to neutrinos (even though the 3D matter field is not directly observable, as we will discuss in Section \ref{sec:results}.).

Lastly, in Ref. \citep{PhysRevD.102.103506}, a distinct but similar technique called Wavelet Phase Harmonics (WPH) was applied on the density field from the {\it Quijote} simulations, but only on 2D projections of it, without including neutrinos. Even though the authors did find an improvement in the forecasted errors over the 2D power spectrum and bispectrum, direct comparisons cannot be made. 

\section{Conclusions}

In this paper, we propose the use of a novel statistic in the context of the galaxy density field, the Wavelet Scattering Transform, in order to optimally leverage the cosmological information content encoded in the 3-dimensional matter distribution of the LSS. 

First conceived of in the context of image processing, and later applied to other (astro-)physical systems, the WST subjects an input field to a series of consecutive non-linear transformations that capture non-Gaussian information, usually sought for with the $n$-point correlation function. The substantial amount of non-Gaussian information in the LSS, that has escaped towards higher order statistics, thus makes such an estimator an ideal candidate for cosmological applications.

In order to demonstrate the WST's ability to optimally exploit the information carried in 3D cosmological density fields, we utilize the publicly available collection of the {\it Quijote} N-body simulations, out of which we extract the Fisher information in 6 $\Lambda$CDM cosmological parameters. After determining the optimal parameters of the WST, we proceed to compare its performance against the conventional power spectrum, as well as another known estimator, the marked power spectrum. 

Through the interplay between several key properties that we identify in this work, the WST is found to deliver a large improvement in the marginalized errors on all parameters, ranging between $1.2-4\times$ tighter than the ones from the  CDM + baryon $P(k)$. We note that, since galaxies are biased tracers of the `cb' field \citep{DESJACQUES20181}, this is realistically the most meaningful one to consider \citep{Villaescusa_Navarro_2014,Castorina_2014,Bayer:2021kwg}.

There are several potential lines of improvement that can bring our analysis closer to realistic observations of the LSS. First of all, the galaxies observed by surveys do not perfectly trace the underlying matter density field, but are \textit{biased} tracers of it \citep{DESJACQUES20181}, residing within dark matter halos. Marginalizing over the unknown bias model in this case may weaken the constraints derived on cosmological parameters, particularly when it comes to the neutrino masses, as discussed in \citep{Bayer:2021kwg}. Secondly, spectroscopic observations are subject to Redshift Space Distortions (RSD) \citep{Hamilton1998}, which introduce an anisotropy to the observed clustering pattern, that needs to be accounted for. In systems with an explicit directional dependence, as is the case with RSD, one can also design wavelets that fully leverage this property\footnote{For example \url{https://github.com/andrew-saydjari/DHC}}. Furthermore, observational effects such as realistic survey geometry and supersample covariance, as well as uncertainties related to galaxy physics that need to be marginalized over \citep{Hahn_2020,Hahn_2021}, are also factors we need to account for before an application to observational data is finally possible. We plan to tackle all of these issues in upcoming work. 

By efficiently combining a set of well-understood mathematical properties, the WST can reliably capture non-Gaussian information way beyond the traditional matter power spectrum. Through the first application on 3D matter overdensity fields in the literature, we demonstrate the great promise it holds in the context of precision cosmology, and pave the way for its future application on observational data. 

\begin{acknowledgments}
GV and CD were partially supported by NSF grant AST-1813694.
This work is supported by the National Science Foundation under Cooperative Agreement PHY-2019786 (The NSF AI Institute for Artificial Intelligence and Fundamental Interactions).
We would like to thank Erwan Allys, Maya Burhanpurkar, Daniel Eisenstein, Doug Finkbeiner, Core Francisco Park, Andrew Saydjari, and Francisco Villaescusa-Navarro for useful comments and discussions.
\end{acknowledgments}

\clearpage
\appendix
\section{Marked Power Spectrum}\label{sec:markedPk}

In this appendix we summarize the basic properties of the marked power spectrum, the predictions of which we used as a comparison. 

In the configuration space, the marked correlation function \citep{https://doi.org/10.1002/mana.19841160115} is a generalization of the standard 2-point correlation function in which each correlated particle (or tracer) is weighted by a function or ``mark'', $m$, that emphasizes a given property of interest. If $\xi(r)$ is the 2-point correlation function at separation $r$, and $W(r)$ the corresponding weighted correlation function, then the marked correlation function, $M(r)$, is commonly defined as: 
\begin{equation}\label{markedcor}
\mathcal{M}(r)= \frac{1}{n(r) \bar{m}^2} \sum_{ij} \delta_D(|\mathbf{x}_i-\mathbf{x}_j|-r)m_i m_j= \frac{1+W(r)}{1+\xi(r)},
\end{equation}
where in Eq.~(\ref{markedcor}) $n(r)$ is the number of pairs of galaxies correlated at a separation $r$, $\bar{m}$ is the mean value of the mark evaluated over the whole sample and $\delta_D$ the Dirac function. 

Marked correlation functions have been utilized to study how the LSS clustering depends on various galaxy properties \citep{Beisbart_2000,
Sheth:2005aj,10.1111/j.1365-2966.2006.10196.x,markeref,10.1111/j.1365-2966.2005.09609.x} through the use of a suitably chosen weight. More recently, an inverse density-weighted mark of the form 
\begin{equation}\label{markdel}
m\left[\mathbf{x},R,\delta_s,p\right]=\left(\frac{1+\delta_s}{1+\delta_s+\delta_R(\mathbf{x})}\right)^p
\end{equation}
was proposed with the aim of testing modified gravity (MG) theories \citep{White_2016}, where $\delta_R$ is the fractional matter overdensity $\delta_m$, smoothed over some scale $R$, and $\delta_s, p $ are two additional free parameters. Choosing values of $p>0$ highlights the low-density regions of the universe, breaking degeneracies between the predictions made by MG theories and General Relativity, as was found to be the case using both dark matter simulations \citep{PhysRevD.97.023535} and also halo catalogues and galaxy mocks \citep{Alam:2020jdv}.

The function of the form (\ref{markdel}) was subsequently successfully applied in the context of constraining neutrino masses by \citep{PhysRevLett.126.011301}, using the Fourier space counterpart of (\ref{markedcor}), the marked power spectrum $M(k)$\footnote{We clarify that, unlike in the expression (\ref{markedcor}), \citep{PhysRevLett.126.011301} did not divide out the clustering contribution $1+\xi(r)$.}. Through a Fisher forecast on the {\it Quijote} simulations \citep{Villaescusa_Navarro_2020}, the combination of parameter values $p=2$, $\delta_s=0.25$ and $R=10$ Mpc/h was found to be the one that minimized the marginalized 1$\sigma$ error on the sum and neutrino masses, which are the values we adopt in our analysis. We evaluate $M(k)$ (and also $P(k)$) up to a $k_{max}=0.58$ h/Mpc, using 91 k-bins, in order to match the minimum physical scale probed by the WST coefficients (whereas \citep{Villaescusa_Navarro_2020,PhysRevLett.126.011301} used a slightly different value of $k_{max}=0.5$ h/Mpc). Finally, we note that \citep{PhysRevLett.126.011301} also considered combinations of marked power spectra with different parameter values, a choice which was shown to produce even tighter constraints to the cosmological parameters. In a similar fashion, combinations of the WST evaluated with different parameters could lead to similar improvements. However, in this work we did not choose to investigate this possibility.

\end{document}